\documentclass[aps,prl,twocolumn,groupedaddress,showpacs]{revtex4}

\usepackage{amsmath}

\begin{document}



\title{Higgs Algebraic Symmetry in Two-Dimensional Dirac Equation}

\author{Fu-Lin Zhang}
\email[Email:]{flzhang@mail.nankai.edu.cn} \affiliation{Theoretical
Physics Division, Chern Institute of Mathematics, Nankai University,
Tianjin 300071, People's Republic of China}

\author{Bo Fu}
\affiliation{Theoretical Physics Division, Chern Institute of
Mathematics, Nankai University, Tianjin 300071, People's Republic of
China}

\author{Jing-Ling Chen}
\email[Email:]{chenjl@nankai.edu.cn}

\affiliation{Theoretical Physics Division, Chern Institute of
Mathematics, Nankai University, Tianjin 300071, People's Republic of
China}

\date{\today}

\begin{abstract}
The dynamical symmetry algebra of the two-dimensional Dirac
Hamiltonian with equal scalar and vector Smorodinsky-Winternitz
potentials is constructed. It is the Higgs algebra, a cubic
polynomial generalization of SU(2). With the help of the Casimir
operators, the energy levels are derived algebraically.
\end{abstract}

\pacs{03.65.Pm; 02.20.-a; 21.10.Sf; 03.65.-w}

\maketitle

The concept of dynamical symmetry (DS) is essential and prevalent
both in classical and quantum mechanics \cite{Greiner}. Hydrogen
atom and isotropic harmonic oscillator are two relatively simple
model with DS, whose classical orbits of motion are closed
\cite{Bertand}. For simplicity, we only concern with the bound sates
of the Coulomb problem in this report. In addition to the orbit
angular momentum corresponding to the rotational symmetry, there
exist more constants of motion in these systems. They have been
proved to be the Rung-Lenz vector \cite{Lenz,Pauli} and the second
order tensors \cite{Frakdin}, in the hydrogen atom and harmonic
oscillator respectively. The algebraic relations of these conserved
quantities reveal the SO(4) symmetry in the hydrogen atom, and SU(3)
in harmonic oscillator. They are called DSs, because the nature of
them
 are not geometrical but the symmetries in the phase space. These symmetries lead to an algebraic approach to determine the
energy levels. Generally, the N-dimensional (ND) hydrogen atom has
the SO(N+1) and the oscillator has the SU(N) symmetry.

 Starting from the feature of the classical orbits, Higgs \cite{Higgs} introduced a generalization of the hydrogen
atom and harmonic oscillator in a spherical space. The conserved
quantities of them construct a cubic polynomial generalization of
SU(2), which is called the Higgs algebra now. Its increasing
applications have been the focus of very active research in recent
year \cite{KK,JLChen,Coherent}. Especially, Floreanini and his
colleagues \cite{CS} find the DS of the two-body Calogero model can
be described by the Higgs algebra. Under an orthogonal
transformation (Eq. (11) in \cite{CS}), their Hamiltonian of the
two-body Calogero model is equivalent to the 2D
Smorodinsky-Winternitz (SW) system \cite{sw1965,sw1994,sw2008}. This
indicates that the DSs, especially which are described by the
polynomial Lie algebra, exist not only in the quantum mechanics
systems with rotation symmetry but also in some non-central
superintegrable potentials \cite{Our}. Moreover, the two-body
Calogero model is shown related to the concept of hidden nonlinear
supersymmetry \cite{NLSS5,NLSS2}. This nonlinear generalization of
supersymmetry is investigated in many systems in recent years
\cite{NLSS3,NLSS1,NLSS7,NLSS6,NLSS8,NLSS4,NLSS9}.

In the relativistic quantum mechanics, the motion of spin-$1/2$
particle satisfies the Dirac equation, which predicts the  intrinsic
magnet moment naturally. The spin-orbit coupling leads to the
breaking of DSs in the Dirac hydrogen atom \cite{Greiner1} and the
Dirac oscillator \cite{DiracOsc}.

Recently, in his illuminating work \cite{U31,U32}, Ginocchio has
found the U(3) and pseudo-U(3) symmetry in the Dirac equation with
scalar and vector harmonic oscillator potentials of equal magnitude.
The Dirac Hamiltonian, with scalar and vector potentials of equal
magnitude (SVPEM), is said to have the spin or pseudospin symmetry
corresponding to the same or opposite sign \cite{ReSymm}. In the
spherical potentials, the total angular momentum can be divided into
conserved orbital and spin parts , which form the SU(2) algebra
separately. Take the spin symmetry case as an example, the two
conserved parts are given by \cite{LS}
\begin{eqnarray}\label{L&S}
\ \vec{L}=\begin{bmatrix}
 \vec{l}&0\\
 0& U_p\vec{l}U_p^{\dag}
 \end{bmatrix},
\ \vec{S}=\begin{bmatrix}
 \vec{s}&0\\
 0& U_p\vec{s}U_p^{\dag}
 \end{bmatrix},
 \end{eqnarray}
where $\vec{l}=\vec{r} \times \vec{p}$,
$\vec{s}=\frac{\vec{\sigma}}{2}$ are the usual spin generators,
$\vec{\sigma}$ are the Pauli matrices, and
$U_p=U_p^{\dag}=\frac{\vec{\sigma}\cdot\vec{p}}{p}$ is the helicity
unitary operator \cite{Hel}. The sum of them equals to the total
angular momentum on account of
$U_p\vec{l}U_p^{\dag}+U_p\vec{s}U_p^{\dag}=\vec{l}+\vec{s}$.
Ginocchio has proved the conserved orbit momentum $\vec{L}$ in Eq.
(\ref{L&S}) to be three of the eight generators of the SU(3)
symmetry group. And, the spin part has no influence on the
Hamiltonian, which behaves like the spin in the nonrelativistic
harmonic oscilltor. We have applied Ginocchio's approach in
\cite{U32} to study the Coulomb potential problem, and found the
SO(4) DS in the Dirac hydrogen atom with spin symmetry \cite{SO4}. A
2D version of this approach also has been introduced to investigate
the SU(2) DS in the 2D Dirac equations with equal scalar and vector
oscillator potentials, and SO(3) for the Coulomb case \cite{2D}.

The Dirac Hamiltonian with SVPEM are derived from the investigation
of the dynamics between a quark and an antiquark
\cite{observe,observe2,quark,LS,Crater}. Many researches about this
type Dirac equations are reported in recent years
\cite{ReSymm,Coulomb,Origin,Antinucleon,Hidden,Pseudospin,equivalent,equivalent2}.
The very lately studies \cite{equivalent,equivalent2} have revealed
that, the motion of a spin-1/2 particle with SVPEM satisfies the
same differential equation and has the same energy spectrum as a
scalar particle. Furthermore, the relativistic energy spectra of the
Klein-Gordon equation with SVPEM are shown to have a one-to-one
relationship with its nonrelativistic limits \cite{KG}. It is worth
while to note that, these results and the concept of the spin or
pseudospin symmetry are
 independent of the shape of the potentials: radial or non-central.
 This leads us to foretell that the Dirac equation with SVPEM have
 the same DS with its nonrelativistic limit, no matter its potential
 is spherical or not. In other words, the conservation of the deformed orbit momentum
 $\vec{L}$ in Eq. (\ref{L&S})
 is not necessary for the presence of DS in the the Dirac Hamiltonian with SVPEM, although it is the first
 step in Ginocchio's approach to deal with the harmonic oscillator \cite{U32} and the Coulomb potentials \cite{SO4}.

As the first trial of our conjecture above, we consider a 2D Dirac
system with equal scalar and vector potentials (ESVP) for
simplicity. Comparing the results in \cite{2D} with \cite{U32,SO4},
one can find the correspondence between the 2D and 3D cases is
straightforward. The non-central potential we choosing is the SW
potential mentioned above,
$V_{sw}(\vec{r})=\frac{1}{2}(x_1^2+x_2^2+k/x_2^2)$ with $k>0$. The
2D Dirac Hamiltonian with ESVP, in the relativistic units,
$\hbar=c=m=1$, takes the form
\begin{eqnarray}\label{H}
H=\vec{\alpha}\cdot\vec{p}+\beta +(1+\beta)\frac{1}{2}V(\vec{r}),
\end{eqnarray}
where $\vec{\alpha}=(\sigma_1, \sigma_2)$ and $\beta=\sigma_3$ are
the Pauli matrices. As shown in \cite{2D}, when the potential
$V(\vec{r})$ in Eq. (\ref{H}) is radial, $H$ commutes with a 2D
version definition of the conserved orbital angular momentum as
\begin{eqnarray}\label{L2}
\ L=\begin{bmatrix}
 \l&0\\
 0& B^{\dag}\frac{l}{p^2}B
 \end{bmatrix},
\end{eqnarray}
where $B=p_1-ip_2$, $B^{\dag}=p_1+ip_2$, and $l=x_1 p_2-x_2 p_1$ is
the usual orbital angular momentum. To derive other additional
conserved quantities in the Coulomb and harmonic oscillator
potentials in \cite{2D}, we assume the constants of motion take the
form as
\begin{eqnarray}\label{Q}
\ Q=\begin{bmatrix}
 Q_{11}&Q_{12}B\\
 B^{\dag}Q_{21}& B^{\dag}Q_{22}B
 \end{bmatrix}.
\end{eqnarray}
The commutation relation $[Q,H]=0$ requires the matrix elements must
satisfy the equations:
\begin{eqnarray}\label{Qeqn}
Q_{12} &=& Q_{21},\nonumber\\
\lbrack Q_{11},V(\vec{r})\rbrack +\lbrack Q_{12},p^2 \rbrack   &=& 0,\\
\lbrack Q_{12},V(\vec{r})\rbrack +\lbrack Q_{22},p^2 \rbrack   &=& 0,\nonumber\\
Q_{11} &=& Q_{12}(2+V(\vec{r}))+Q_{22}p^2.\nonumber
\end{eqnarray}
They are the same as the 3D case \cite{U32}.

When the potential in Eq. (\ref{H}) takes the SW form,
$V(\vec{r})=V_{sw}(\vec{r})$, the deformed orbit momentum $L$ in Eq.
(\ref{L2}) is no longer a constant of motion. But, one can notice
the derivative process, from the ansatz form of $Q$ to the
conditions of its element in Eq. (\ref{Qeqn}), is not relying on the
form of $V(\vec{r})$. In addition, $L$ also satisfies the conditions
when the potential $V(\vec{r})$ is radial symmetric. Therefore, we
presume all the generators of the symmetry algebra of the
Hamiltonian $H$ with the SW potential can be determined from the
conditions.


Three solutions of Eq. (\ref{Qeqn}) with
$V(\vec{r})=V_{sw}(\vec{r})$ are found, whose independent elements
are given by
\begin{eqnarray}\label{DDQ}
D_1:&&D^{(1)}_{12}=x_1^2\bigr(x_2^2-\frac{k}{x_2^2}\bigr)\bigr[2+V_{sw}(\vec{r})\bigr]-2l^2+2p_1^2\frac{k}{x_2^2} \nonumber\\
&&\ \ \ \ \ \ \ \ \ \ +2x_1x_2p_1p_2+2p_1p_2x_1x_2,  \nonumber\\
&&D^{(1)}_{22}=x_1^2\bigr(x_2^2-\frac{k}{x_2^2}\bigr)+\frac{4p_1^2p_2^2}{p^2}; \nonumber\\
D_2:&&D^{(2)}_{12}=x_1^2(x_2p_2+p_2x_2)-\bigr(x_2^2-\frac{k}{x_2^2}\bigr)(x_1p_1+p_1x_1),  \nonumber\\
&&D^{(2)}_{22}=\frac{2}{p^2}\bigr[p_2^2(x_1p_1+p_1x_1)-p_1^2 (x_2p_2+p_2x_2)  \bigr];  \nonumber\\
Q_3:&&Q^{(3)}_{12}=\frac{1}{2}\bigr(x_1^2-x_2^2-\frac{k}{x_2^2}\bigr),  \nonumber\\
&&Q^{(3)}_{22}=\frac{p_1^2-p_2^2}{p^2}. \nonumber
\end{eqnarray}
The other elements, $D^{(1)}_{i 1}$, $D^{(2)}_{i 1}$ and $Q^{(3)}_{i
1}$ ($i=1,2$), can be obtained easily from the first and the last
relations in Eq. (\ref{Qeqn}). They combine into the three constants
of motion in the form of Eq. (\ref{Q}) as
\begin{eqnarray}\label{T}
\ T_{i}=\begin{bmatrix}
 T^{(i)}_{11}&T^{(i)}_{12}B\\
 B^{\dag}T^{(i)}_{21}& B^{\dag}T^{(i)}_{22}B
 \end{bmatrix}, \nonumber
\end{eqnarray}
with $T=D$ for $i=1,2$ and $T=Q$ for $i=3$.
We can define the normalized generators, $D_{\pm}=D_1 \pm i
\sqrt{\mathcal{G}} D_2$ and $D_{3}=[4\sqrt{\mathcal{G}}]^{-1} Q_3 $,
with $\mathcal{G}=2(H+1)$ being a constant for a fixed energy level.
They satisfy the Higgs algebra relations
\begin{eqnarray}\label{Higgs}
\lbrack D_{3},D_{\pm} \rbrack &=& \pm D_{\pm},\\
\lbrack D_{+ },D_{-} \rbrack &=& c_3 D^3_3 + c_1 D_3 +c_0,\nonumber
\end{eqnarray}
where $c_3= -1024 \mathcal{G}^2$, $c_1=64 (\mathcal{F}-2
\mathcal{G}) \mathcal{G} + 32 k \mathcal{G}^3$, $c_0= 8 k
\mathcal{G}^2 \sqrt{(\mathcal{F}+\mathcal{G})\mathcal{G}} $ and
$\mathcal{F} = (H^2-1)^2-\mathcal{G}$. The Casimir operator of the
Higgs algebra can be obtained immediately \cite{CofH,XJZ}
\begin{eqnarray}\label{Casimir}
\mathcal{C} &=& \{ D_{+}, D_{-} \} +\frac{c_3}{2} D^4_3 +(c_1 +
\frac{c_3}{2})D^2_3 +2 c_0  D_3 \nonumber\\
&=& 2 \mathcal{F}(\mathcal{F}-8 \mathcal{G})- 2 k^2 \mathcal{G}
(\mathcal{F}+ 4 \mathcal{G}).
\end{eqnarray}

%
%


In the basis of the common eigenstates of $H$ and $D_3$, $ H |E,m
\rangle = E  |E,m \rangle$, $ D_3 |E,m \rangle = m|E,m \rangle$, $
D_{\pm}$ are the ladder operators of $D_3$ ,  $D_3 D_{\pm} |E,m
\rangle =(m \pm 1 )D_{\pm} |E,m \rangle$. There exist a highest and
a lowest weights for a fixed energy level, denoted as  $D_{+} |E,
\overline{m} \rangle=0 $ and $D_{-} |E, \underline{m} \rangle=0 $.
Let $S_{+}=2D_{-}D_{+}$ and $S_{-}=2D_{+}D_{-}$, one can obtain from
the relations in Eq. (\ref{Higgs}) and (\ref{Casimir})
\begin{eqnarray}\label{Spm}
S_{\pm}= \mathcal{C}- \bigr[ \frac{c_3}{2} D^2_3 (D_3 \pm 1) ^2 +
c_1 D_3 (D_3 \pm 1 )+ c_0 (2D_3 \pm 1)   \bigr]. \nonumber
\end{eqnarray}
Operating with them on $|E,\overline{m} \rangle$  and
$|E,\underline{m} \rangle$ respectively, we obtain
\begin{eqnarray}\label{mpm}
\overline{m} &=& \frac{1}{8}\bigr[ -4 - \sqrt{4+8k (E+1)} +
\sqrt{2(E+1)} (E-1 ) \bigr], \nonumber \\
\underline{m} &=& \frac{1}{8} \bigr[ \lambda- \sqrt{2(E+1)} (E-1)
\bigr],
\end{eqnarray}
where $\lambda=2$ or $6$. The degeneracy of states in a given energy
level should be a natural number, which leads to $\overline{m} -
\underline{m} = n = 0 , 1 ,2 ...$ Substituting it into Eq.
(\ref{mpm}), we find the eigenvalues of $H$ satisfy the equation
\begin{eqnarray}\label{energy}
\sqrt{\frac{1}{2}(E+1)}(E-1)=N+\frac{3}{2}+\sqrt{\frac{1}{4}+\frac{k}{2}(E+1)},
\end{eqnarray}
where $N=2n$ or $2n+1$ corresponding to the different values of
$\lambda$ in Eq. (\ref{mpm}). The degeneracy of the energy level is
the same as the two-body Calogero model given in \cite{CS},  $d= n+1
=[N/2]+1$, $[x]$ being the integer part of $x$.

When the parameter $k \rightarrow 0$, the energy levels became the
results in the 2D harmonic oscillator \cite{2D}, but different in
the value range of the total quantum number. The difference is
derived from the fact that the limit of the SW potential is a
harmonic oscillator potential with a infinite barrier along the
$x_1$ axis. When $E \rightarrow 1$, the nonrelativistic limit of the
energy levels is given by $E-1= N+ 3/2 + \sqrt{k+1/4}$, which agrees
with the nonrelativistic results \cite{sw1994}.

In summary, we have shown that the 2D Dirac system with equal scalar
and vector SW potentials has a DS described by the Higgs algebra.
The three generators are derived by using a 2D version of the ansatz
form given by Ginocchio. The relation of the Casimir operator of the
Higgs algebra and the Hamiltonian leads to an algebraic solution of
the relativistic energy spectrum.

This is our first attempt to investigate the DS in noncentral Dirac
system. And as we know, it is also the first example in the Dirac
quantum mechanics, whose dynamical symmetry is described by the
Higgs algebra. Actually, the 3D Dirac equations with spin symmetry
or pseudospin symmetry are more real, which exist frequently in
antinucleon and nucleon spectra \cite{ReSymm}. Whereas, it is
natural to generalize our treatment to the 3D case to study the
dynamical symmetries in some noncentral potentials, such as the
anisotropic harmonic oscillator with rational frequency ratio and
the caged anisotropic oscillator potentials \cite{sw2008}.

\begin{acknowledgments}
 This work is supported in part by NSF of China (Grants No. 10975075) and Program for New Century Excellent Talents in
University. The Project-sponsored by SRF for ROCS, SEM.
\end{acknowledgments}

\bibliography{HiggsDirac}

\end{document}